\documentclass[aps,prx,reprint,superscriptaddress,amsmath,amssymb]{revtex4-2}
\usepackage{graphicx}
\graphicspath{{_Figures/}{./}}
\usepackage{hyperref}
\usepackage{bm}
\usepackage{booktabs}
\usepackage{tabularx}
\usepackage{array}
\usepackage{threeparttable}
\usepackage{makecell}
\usepackage{ragged2e}
\usepackage{setspace} % line spacing controls
\usepackage{float} % allow [H] float placement

\renewcommand{\arraystretch}{1.2}
\usepackage{array}
\newcolumntype{L}[1]{>{\raggedright\arraybackslash}p{#1}}
\newenvironment{aidisclosure}{\par\vspace{1em}\noindent\textbf{AI Disclosure:} }{\par\vspace{1em}}
\begin{document}
\onehalfspacing

\title{Microstructural Topology as a Prescriptor for Quantum Coherence: \\
\textit{Towards A Unified Framework for Decoherence in Superconducting Qubits}}

\author{Vinayak P. Dravid}
\email{v-dravid@northwestern.edu}
\affiliation{Department of Materials Science \& Engineering,
  Northwestern University, Evanston, IL~60208}
\affiliation{Northwestern University Atomic and Nanoscale
  Characterization Experimental Center (NUANCE),
  Evanston, IL~60208}
\affiliation{Applied Physics Graduate Program,
  Northwestern University, Evanston, IL~60208}
\affiliation{Department of Chemistry (courtesy),
  Northwestern University, Evanston, IL~60208}

\author{Akshay A. Murthy}
\affiliation{Superconducting Quantum Materials and Systems Center
  (SQMS), Fermi National Accelerator Laboratory,
  Batavia, IL~60510}

\author{Peter Lim}
\affiliation{Applied Physics Graduate Program,
  Northwestern University, Evanston, IL~60208}

\author{Gabriel T. dos Santos}
\affiliation{Department of Materials Science \& Engineering,
  Northwestern University, Evanston, IL~60208}

\author{Ramandeep Mandia}
\affiliation{Department of Materials Science \& Engineering,
  Northwestern University, Evanston, IL~60208}

\author{James M. Rondinelli}
\affiliation{Department of Materials Science \& Engineering,
  Northwestern University, Evanston, IL~60208}
\affiliation{Applied Physics Graduate Program,
  Northwestern University, Evanston, IL~60208}

\author{Mark C. Hersam}
\affiliation{Department of Materials Science \& Engineering,
  Northwestern University, Evanston, IL~60208}
\affiliation{Applied Physics Graduate Program,
  Northwestern University, Evanston, IL~60208}
\affiliation{Department of Chemistry (courtesy),
  Northwestern University, Evanston, IL~60208}

\author{Roberto dos Reis}
\affiliation{Department of Materials Science \& Engineering,
  Northwestern University, Evanston, IL~60208}
\affiliation{Northwestern University Atomic and Nanoscale
  Characterization Experimental Center (NUANCE),
  Evanston, IL~60208}
\affiliation{Applied Physics Graduate Program,
  Northwestern University, Evanston, IL~60208}

\date{March 2026}

%% ============================================================
\begin{abstract}
In superconducting quantum circuits, decoherence improvements are frequently obtained through process interventions that simultaneously modify surface chemistry, microstructural topology, and device geometry, leaving mechanistic attribution structurally underdetermined. Predictive materials engineering requires measurable structural statistics to be separated from geometry-dependent coupling coefficients into independently testable factors. We introduce the concept of classical and quantum microstructure. In that context,  we formulate a channel-wise separable framework for decoherence in superconducting transmon qubits in which each loss channel~$j$ is described by a reduced prescriptor $\Pi_j = \rho_j G_j$. Here $\rho_j$ is a channel-specific microstructural state variable determined independently of device geometry, and $G_j$ is a geometry-dependent coupling functional computable from field solutions without reference to surface chemistry. We derive this product form from a spatially resolved kernel representation, $\mathcal{O}_j = \int_{\Omega_j}d_j(\mathbf{r})\,K_j(\mathbf{r};\mathcal{G})\, d^n r + \Delta_j$, and establish a perturbative separability criterion that defines the regime where independent variation of $\rho_j$ and $G_j$ is valid. The framework specifies five prescriptor classes for dominant loss pathways in transmon-class devices. Falsifiability is operationalized through a pre-committed $2\!\times\!2$ experimental protocol in which $\rho_j$ and $G_j$ must satisfy independent ratio checks within propagated uncertainty. A Minimum-Dataset Specification standardizes reporting for cross-laboratory inference. Part~I establishes the conceptual and mathematical architecture; coordinated experimental validation is reserved for Part~II.
\end{abstract}

\maketitle

%% ============================================================
\section{Introduction}
\label{sec:intro}

Materials science becomes predictive when empirical correlations are reorganized into structure-property relations that isolate a measurable structural variable from a geometry- or loading-dependent coupling coefficient. The Hall--Petch relation~\cite{hall1951,petch1953} established that polycrystalline yield strength scales as $\sigma_y = \sigma_0 + k_y d^{-1/2}$, identifying grain-boundary density as the operative intensive variable and $k_y$ as a geometry-independent coupling constant.  The Griffith--Irwin fracture framework~\cite{irwin1957} separated flaw-population statistics from the stress-intensity factor, transforming empirical fracture observation into predictive engineering.  In both cases the lasting contribution was not a new microscopic mechanism but a disciplined reduction: the identification of which structural statistic must be measured independently and which geometry factor must be computed independently, so that predictions can be made, transferred, and falsified.

At present, improvements in qubit coherence are structurally underdetermined: process modifications simultaneously alter surface chemistry, microstructural topology, and circuit geometry, preventing mechanistic attribution even in controlled experiments. The logic of this paper follows the same two-step pattern. First, we identify that decoherence in superconducting qubits is currently \textit{attribution-underdetermined}: successful process interventions routinely alter surface chemistry, microstructural topology, and circuit geometry simultaneously, so no single experiment can isolate the operative variable. Second, we propose a disciplined separation. We define a \textit{prescriptor} as a falsifiable structure–property relation that separates (i) a microstructural state variable measurable independent of device geometry, and (ii) a geometry-dependent coupling functional computable independent of materials chemistry. A descriptor correlates retrospectively. A prescriptor predicts under pre-committed experimental tests.

The five prescriptor channels (I--V) developed in this paper are specific instantiations of this core concept. The mathematics that follows formalizes this separation; readers who prefer to anchor the formalism in physical motivation may find it useful to read Sec.~\ref{sec:channels} in parallel with Sec.~\ref{sec:architecture}.

\begin{figure}[t]
  \centering
  \includegraphics[width=\linewidth]{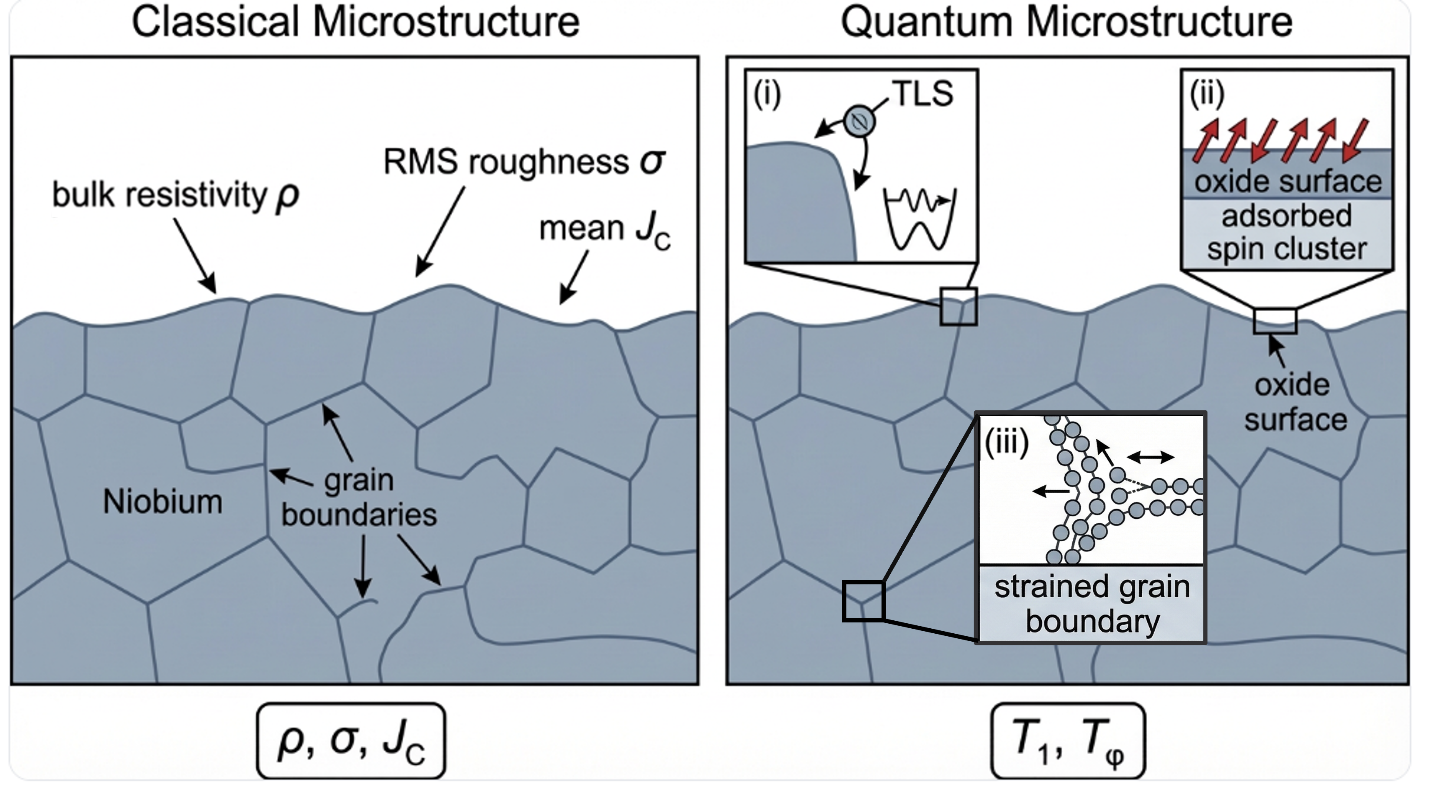}
  \caption{\textbf{Classical versus quantum microstructure.} (a)~Classical microstructure: ensemble-averaged structural features govern amplitude-based observables such as bulk resistivity~$\rho$, RMS roughness~$\sigma$, and mean critical current density~$J_c$. (b)~Quantum microstructure: coherence is governed by a statistically sparse population of structurally severe defect configurations - atomic edge cusps hosting two-level systems~(TLS), strained grain boundaries, adsorbed spin clusters - whose coupling to the qubit mode is superlinear in local structural severity. These rare, strongly coupled sites motivate distribution-resolved descriptors such as the curvature second moment~$\mu_2$, which may correlate more strongly with coherence observables than ensemble averages.  This distinction determines which statistics are predictive, not merely descriptive.}
  \label{fig:microstructure}
\end{figure}

Early experiments resolving discrete two-level system (TLS) where qubit coupling established the microscopic origin of dielectric loss in Josephson-junction circuits~\cite{simmonds2004,martinis2005,cooper2004}. Those measurements identified individual atomic-scale defects as the dominant decoherence agents, motivating two decades of surface and interface engineering~\cite{muller2019,lisenfeld2019}. The present framework builds directly on that lineage but addresses a different level of description: where those experiments established the microscopic origin of TLS-mediated loss at the atomic scale, the prescriptor framework addresses the mesoscopic materials question, which measurable microstructural statistic rationalizes the density and severity of TLS-hosting sites across device geometries and fabrication processes, and how to test that prediction independently. Superconducting qubits present a closely related but unresolved inference problem.  Coherence times have advanced from submicrosecond values to the millisecond regime~\cite{wolff2026structural,oliver2013,krantz2019,kjaergaard2020,place2021} through improvements in materials processing, surface treatment, and circuit design.  Yet many successful interventions alter surface chemistry, microstructural topology, and circuit geometry simultaneously, so measured improvements in $T_1$ or $T_\varphi$ cannot be uniquely attributed to any single mechanism. This ambiguity has both practical and structural dimensions. Without a formalism that separates microstructural statistics from geometry-dependent coupling, no single-process-change experiment can falsify a mechanistic claim about decoherence origin, because every relevant variable is perturbed at once~\cite{wenner2011,wang2015,woods2019,altoe2022,verjauw2021}.

A central conceptual distinction motivates the framework developed here: the difference between \textit{classical microstructure} and \textit{quantum microstructure} (FIG.~\ref{fig:microstructure}). Classical microstructure is adequately characterized by ensemble averages because amplitude-based observables (resistivity, thermal conductivity, critical current density) respond to broad defect populations. Quantum coherence is qualitatively different: rare, spatially
localized, and structurally severe configurations can dominate the decoherence budget because their coupling to the qubit mode scales superlinearly with structural severity.  In this regime, descriptors based on ensemble averages (e.g., RMS roughness) are insufficient. Distribution-resolved statistics, moments of curvature, tail fractions of spin-cluster density, local severity of oxide stoichiometry, therefore become the physically relevant candidates for a predictive decoherence framework.

\begin{figure*}[t]
  \centering
  \includegraphics[width=\linewidth]{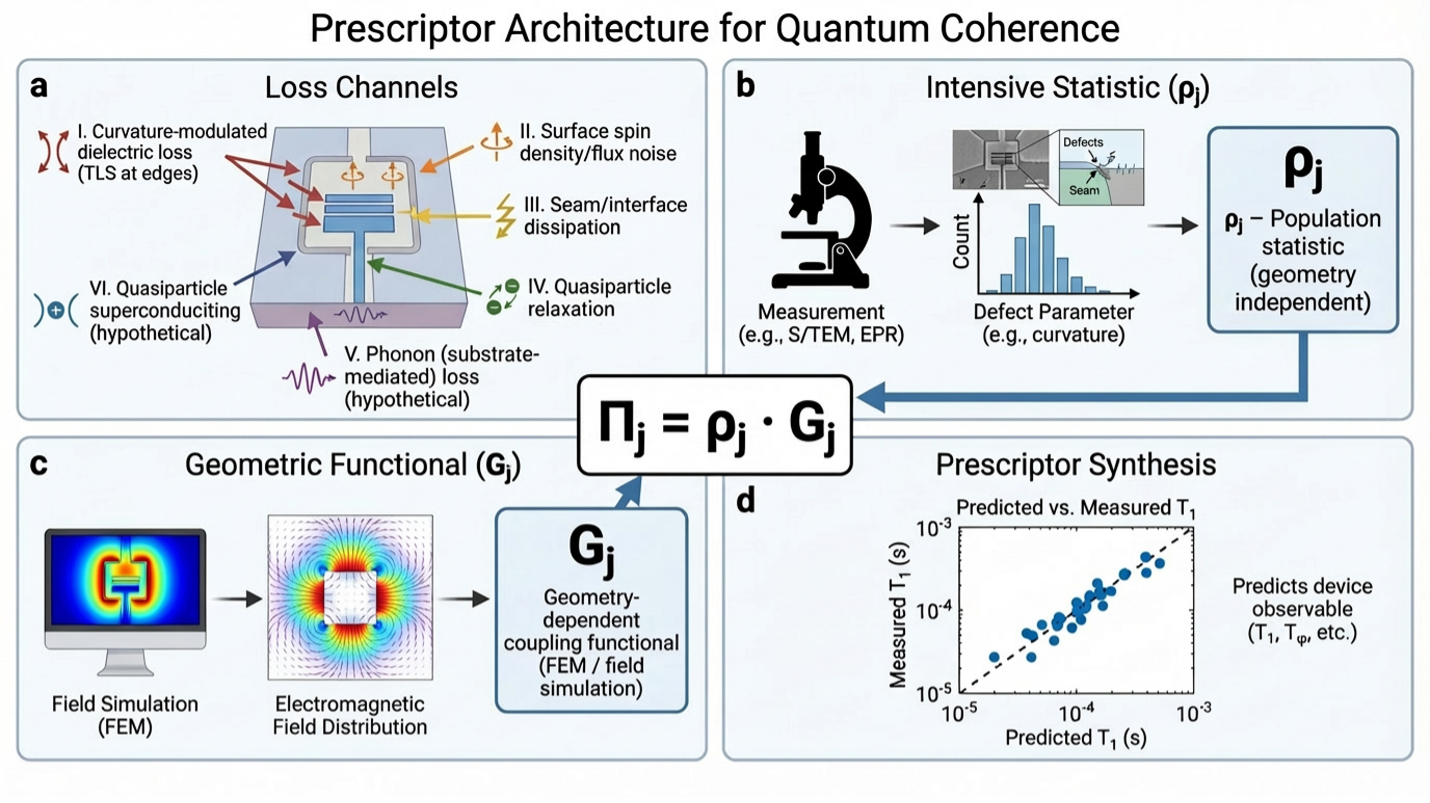}
  \caption{\textbf{Prescriptor framework overview.} (a)~Physical transmon device with five representative loss channels. (b)~Extraction of channel-specific microstructural state variables $\rho_j$ from witness-sample measurements, independent of device geometry. (c)~Computation of geometry-dependent coupling functionals $G_j$ from finite-element field solutions, independent of surface chemistry. (d)~Framework illustration and Part II validation target: comparing the predicted scalar prescriptor $\Pi_j = \rho_j G_j$ against independently measured observables. The vertical division between panels~(b) and~(c) represents the structural independence of $\rho_j$ and $G_j$, validated experimentally via the $2\!\times\!2$ protocol of Sec.~\ref{sec:falsifiability}.}
  \label{fig:overview}
\end{figure*}

We therefore formulate a \textit{channel-wise separable reduced-order framework} for decoherence in superconducting transmon qubits. The central claim is not that every decoherence mechanism is universally multiplicative in a product of density and coupling, but that several important channels admit a useful separable approximation when disorder is dilute, backaction on the mode is weak, and geometry-induced redistribution of the disorder ensemble is perturbatively small. We identify the experimentally relevant regimes in which leading-order factorization is controlled and testable. In that regime, one may define a microstructural state variable from witness-sample measurements and a geometry-dependent coupling functional from field calculations, and their product serves as a testable reduced descriptor. We call this descriptor a \textit{prescriptor} because it is formulated to prescribe a falsifiable structure-property relation for predictive engineering, rather than merely describe retrospective data.

Throughout this paper, \textit{the framework} refers specifically to the separability and falsifiability structure: the claim that $\Pi_j = \rho_j G_j$ is a controlled reduced-order approximation in the dilute-defect, weak-backaction regime, and that this approximation admits a pre-committed experimental test. Each of the five prescriptor classes is a candidate instantiation of that framework, a testable hypothesis about which microstructural statistic serves as $\rho_j$ for a given loss channel. The instantiations are physically motivated, but they are not assertions of the core framework; they stand or fall on their own experimental merits. In the dilute, weak-backaction limit, the total rate reduces to a sum over independent localized contributions, yielding a controlled leading-order product form after coarse-graining. The product form $\Pi_j = \rho_j G_j$ is therefore not introduced as an unrestricted \textit{ansatz}, but as the leading-order result of a short derivation. The total decoherence rate for channel~$j$ is a sum over all defects: $\Gamma_j = (2\pi/\hbar)\sum_d |\langle f | h_d | i \rangle|^2\delta(E_f - E_i)$. In the dilute-defect limit, inter-site correlations are negligible and the sum factorizes into a defect count times a single-defect rate: $\Gamma_j = \rho_j \cdot \gamma_\mathrm{single}(\omega,\mathcal{G})$. Because the matrix element in the weak-coupling limit is determined primarily by the local field amplitude at the defect site, fixed by mode geometry and perturbatively insensitive to defect density, one identifies $\gamma_\mathrm{single} \equiv G_j$, and the leading-order product form follows directly: $\Pi_j = \rho_j G_j$.

The measurement hierarchy used throughout this paper has three tiers. At the \textit{device level} are directly observable quantities ($T_1$, $T_\varphi$, $A_\Phi$) that characterize qubit performance but do not identify independently addressable causes.  At the \textit{prescriptor level} are the upstream variables $\rho_j$ and $G_j$, and their product $\Pi_j = \rho_j G_j$, which predicts the device-level observable via a channel-specific proportionality. At the \textit{experimental-design level} are the falsifiability protocols - the $2\!\times\!2$ decoupling matrix (Sec.~\ref{sec:falsifiability}) and the Minimum-Dataset Specification (Sec.~\ref{sec:mds})- that specify the experimental conditions under which a prescriptor prediction can be tested.  The prescriptor formalism does not replace measurement of $T_1$ and $T_\varphi$; it provides the upstream structure that makes those measurements mechanistically interpretable. As an illustrative example, curvature-controlled dielectric loss can be described by taking the curvature second moment $\mu_2$ as the relevant microstructural statistic and the local electric-field concentration as the coupling functional. Under fixed geometry, this implies that $T_1^{-1}$ scales with $\mu_2$, providing a reliable instance of a distribution-resolved structural statistic entering a testable relation. We define a $2!\times!2$ experimental protocol that forces independent validation or falsification of the framework without parameter fitting.

The paper is organized as follows. Section~\ref{sec:architecture} establishes the spatially resolved kernel representation, derives the reduced prescriptor as a leading-order coarse-graining, states the perturbative separability criterion, and recovers participation-ratio analysis as a special case. Section~\ref{sec:channels} develops the five prescriptor classes in order of decreasing formal maturity. Section~\ref{sec:coupling} provides an operational cross-channel coupling map for transmon-class devices. Section~\ref{sec:falsifiability} formalizes the $2\!\times\!2$ pre-commitment protocol as a quantitative falsifiability test. Section~\ref{sec:mds} defines the Minimum-Dataset Specification. Section~\ref{sec:roadmap} describes the four experimental axes for Part~II validation. Section~\ref{sec:discussion} situates the framework in the context of materials science and quantum engineering.

%% ============================================================
\section{Framework Architecture and Separability Criterion}
\label{sec:architecture}

FIG~\ref{fig:overview} maps the prescriptor architecture; this section formalizes it by defining the geometry-dependent kernels, their scalar reductions, and the separability conditions under which the product form is mathematically justified.

\subsection{Spatially resolved kernel representation}

We begin not with the scalar product form but with a representation that makes explicit the spatial structure of the coupling between disorder and field. For decoherence channel~$j$, let $d_j(\mathbf{r})$ denote a local defect-state field, the effective local disorder density or susceptibility at position~$\mathbf{r}$, and let $K_j(\mathbf{r};\mathcal{G})$ denote the geometry-dependent coupling kernel determined by the device geometry and mode structure~$\mathcal{G}$. The observable associated with channel~$j$ is then written as
\begin{equation}
  \mathcal{O}_j
  = \int_{\Omega_j}
    d_j(\mathbf{r})\,K_j(\mathbf{r};\mathcal{G})\,d^n r
    + \Delta_j,
  \label{eq:kernel}
\end{equation}
where $\Omega_j$ is the spatial domain relevant to channel~$j$, and $\Delta_j$ collects higher-order contributions. Equation~(\ref{eq:kernel}) is the starting spatial representation. In the dilute-defect, weak-backaction limit, replacing the spatial disorder field $d_j(\mathbf{r})$ with its sufficient statistic $\rho_j$ yields the reduced scalar form $\Pi_j = \rho_j G_j$.

This coarse-graining follows directly from treating the channel-$j$ perturbation Hamiltonian as a sum over localized non-interacting contributions, $H'_j = \sum_{d \in j} h_d$. To leading order in Fermi's golden rule, the transition rate is:
\begin{align}
  \Gamma_j
  &= \frac{2\pi}{\hbar}
     \sum_{d \in j}
     \bigl|\langle f | h_d | i \rangle\bigr|^2
     \delta(E_f - E_i),
  \label{eq:goldenrule}
\end{align}
where the sum runs over all defects of class~$j$. If each matrix element is controlled primarily by the local field amplitude at the defect site, the sum becomes a weighted integral over space and Eq.~(\ref{eq:goldenrule}) reduces to the continuum form of Eq.~(\ref{eq:kernel}).

\subsection{Reduced scalar prescriptor}

A reduced scalar descriptor emerges from Eq.~(\ref{eq:kernel}) when the disorder field can be summarized by a scalar state variable $\rho_j = \mathcal{M}_j[d_j]$ and the coupling kernel can be summarized by a scalar functional $G_j = \mathcal{F}_j[K_j]$, where $\mathcal{M}_j$ and $\mathcal{F}_j$ are channel-specific coarse-graining operators.  The leading-order approximation is
\begin{equation}
  \mathcal{O}_j \approx C_j\,\rho_j\,G_j,
  \label{eq:leadingorder}
\end{equation}

where $C_j$ is a channel-dependent scalar prefactor. This prefactor absorbs fundamental physics constants, orientation-averaging coefficients, and necessary dimensional conversions (e.g., $C_\Phi = \mu_B^2/\Phi_0^2$ for flux noise), ensuring that the proportionality correctly maps the prescriptor to the device-level observable.

We define the \textit{prescriptor} (Fig.~\ref{fig:overview}(d)) as:
\begin{equation}
  \boxed{\Pi_j \equiv \rho_j\,G_j,}
  \label{eq:prescriptor}
\end{equation}
where the proportionality $\mathcal{O}_j \approx C_j\Pi_j$ maps the prescriptor to the device-level observable. In the limit of spatially uniform disorder, the framework reduces exactly to standard participation-ratio formulations, establishing it as a strict generalization rather than an alternative description.

The present reduced-order form does not replace the standard Hamiltonian or noise-spectral description of qubit decoherence. Rather, it supplies a materials-facing coarse-graining of that description. In the usual formulation, a channel-specific transition or dephasing rate is governed by a bath spectral factor together with a device-dependent matrix element. The novelty is not the density–coupling product itself, but isolating variables that can be measured and computed independently, and pairing them with a pre-committed falsifiability protocol.

Here, the microstructural state variable $\rho_j$ plays the role of the experimentally accessible disorder-side intensity, while the geometry functional $G_j$ captures the mode- and layout-dependent coupling weight obtained from field solutions. The prescriptor $\Pi_j = \rho_j G_j$ is therefore intended as the leading-order materials realization of the same rate structure, written in a form that can be independently measured, computed, and prospectively falsified across fabrication and geometry splits.

Both $\rho_j$ and $G_j$ are positive-definite by construction: $\rho_j$ is a population density (defect density, spin density, conductance) and $G_j$ is a squared-field overlap integral ($|\mathbf{E}|^2$, $|\mathbf{B}|^2$, $|\mathbf{J}_s|^2$), ensuring $\Pi_j \geq 0$ for all channels. The prescriptor captures a loss rate, not a gain; coherence enhancement corresponds to reducing $\Pi_j$ toward zero, not to negative values.  Apparent enhancement mechanisms, such as coherent defect-qubit interactions beyond the perturbative regime, are captured through reductions in effective $\rho_j$ (e.g., passivation reducing defect density) or redistribution of coupling geometry (e.g., mode reshaping reducing $G_j$).

The variable $\rho_j$ is a \textit{channel-specific microstructural state variable} (Fig.~\ref{fig:overview}(b)): it represents the intensity of the disorder ensemble relevant to channel~$j$, in whatever form is appropriate for that channel.  Depending on the channel it may be an areal defect density, a distribution moment, an interfacial material parameter, or an environmental state variable. The unifying requirement is not identical microscopic meaning across channels, but that $\rho_j$ be measurable independently of device geometry to leading order under controlled processing conditions. The variable $G_j$ is a \textit{geometry-dependent coupling functional} (Fig.~\ref{fig:overview}(c)): it encodes how strongly a unit of the relevant disorder couples to the qubit mode, as determined by the mode-field solution.

We emphasize that $\rho_j$ is not required to have a universal microscopic form across all channels. Its role is operational: for each channel, $\rho_j$ denotes the lowest-order independently measurable state variable that summarizes the relevant disorder ensemble for predictive transfer across geometries.

This is analogous to classical structure-property formulations, where grain-boundary density, flaw-population statistics, and interfacial conductance are all admissible intensive variables despite representing distinct microscopic objects. The unifying requirement is separable measurability and predictive utility, not identical microscopic meaning.

The prescriptor framework is defined at the qubit operating temperature ($T \ll T_c$, typically 10-20~mK), where $\rho_j$ values are measured and $G_j$ is computed.  Temperature dependence enters exclusively through $\rho_j$ via thermally activated processes (e.g., quasiparticle density $\bar{n}_{qp}$ in Prescriptor~IV(b) scales exponentially with $T$), while $G_j$ is assumed to be temperature-independent to leading order. This working assumption is justified deep in the superconducting state ($T \ll T_c$), where macroscopic electromagnetic mode volumes and penetration depths are effectively frozen, rendering any temperature-induced shifts in the geometric coupling response perturbatively small. This thermal factorization preserves the separability of materials and geometry degrees of freedom.

The five principal instantiations of $\rho_j$ and $G_j$ used here are summarized below.

\subsection{Perturbative separability criterion}
\label{sec:separability}

Separability is the central assumption of the framework, and its status must be stated precisely and verified experimentally.  Let $\delta\rho_j(\mathrm{geom})$ denote the change in the microstructural state variable induced by a geometry variation at fixed process chemistry, and let $\delta G_j(\mathrm{chem})$ denote the change in the coupling functional induced by a surface-chemistry change at fixed geometry. The reduced-order prescriptor is perturbatively controlled when
\begin{equation}
  \left|\frac{\delta\rho_j(\mathrm{geom})}{\rho_j}\right| \ll 1
  \quad \text{and} \quad
  \left|\frac{\delta G_j(\mathrm{chem})}{G_j}\right| \ll 1.
  \label{eq:separability}
\end{equation}

Equation (\ref{eq:separability}) is not a post hoc condition but a prospective design criterion: \textit{experiments violating these bounds are expected to fail the prescriptor test}. In practice, $\Pi_j = \rho_j G_j$ holds when geometry changes do not materially redistribute the disorder ensemble ($|\delta \rho_{\text{geom}}|/\rho_j \ll 1$), and when chemistry changes do not alter the computed coupling functional ($|\delta G_{\text{chem}}|/G_j \ll 1$). The $2\!\times\!2$ protocol of Sec.~\ref{sec:falsifiability} serves as the experimental test of this criterion.

Three standard approximations underlie this reduced form: (i)~\textit{dilute defects} with negligible spatial correlations; (ii)~\textit{weak backaction} where defect coupling does not reshape the mode; and (iii)~\textit{linear response} of the defect bath. The framework breaks down predictably if defects percolate, TLS saturate, or geometry mechanically redistributes the underlying defect ensemble.

The claim of the present work is therefore deliberately limited. The framework is intentionally restricted to the dilute-defect, weak-backaction regime, where separability is a controlled approximation rather than an assumption. We assert that, in the dilute-defect and weak-backaction regime, a channel-wise separable reduction is the appropriate null model for prospective materials inference. Its value lies in being falsifiable: failure of the row and column ratio checks identifies the onset of cross-perturbation, strong coupling, or missing state variables, and therefore signals when higher-order descriptions are required.

\subsection{Relation to participation-ratio analysis}

Standard participation-ratio analysis writes
$Q^{-1} = \sum_i p_i \tan\delta_i$, assigning a spatially uniform
loss tangent~$\tan\delta_i$ to each interface
region~\cite{wenner2011,wang2015,gao2008,woods2019, crowley2023}.  This expression
is recovered exactly as the spatially homogeneous-$\rho_j$ limit of
Eq.~(\ref{eq:kernel}): when $d_j(\mathbf{r}) = \text{const}$ within
region~$i$, the kernel integral reduces to $\tan\delta_i \times p_i$
and the standard participation-ratio result is reproduced identically.

Participation-ratio analysis is recovered as a degenerate limit of
the prescriptor formalism: the standard expression
$Q^{-1} = \sum_i p_i \tan\delta_i$ corresponds exactly to the
spatially homogeneous-$\rho_j$ limit of Eq.~(\ref{eq:kernel}),
in which the disorder field is treated as uniform within each
interface region.  The prescriptor framework therefore extends, rather
than replaces, participation-ratio analysis: participation ratios
specify \textit{where} in the device loss is geometrically
concentrated; prescriptors specify \textit{which} microstructural
feature drives that loss and \textit{how} to engineer it.

Participation ratios localize where electromagnetic energy resides but do not identify which independently measurable microstructural statistic controls the loss within that region.

The prescriptor framework extends this by assigning a channel-specific state variable that can be measured on witness samples and tested for predictive transfer across geometries.

%% ============================================================
\paragraph{Channel-wise prescriptor summary.}
For the transmon regime, the leading prescriptor classes are:

\begin{itemize}
\item \textbf{I-TLS:} Curvature-conditioned dielectric loss at electrode interfaces, with \( \rho_j = \mu_2 \equiv \langle R_c^{-2}\rangle \) and geometry functional \( G_I \); the leading relation is \( \Pi_I=\mu_2 G_I \), with observable \(Q^{-1}\) or \(T_1\).
\item \textbf{II-Spin:} Surface-spin-driven \(1/f\) flux noise, with \( \rho_j=\rho_{\mathrm{spin}} \) and geometry functional \( G_{\Phi} \); the leading relation is \( \Pi_{\mathrm{spin}}=\rho_{\mathrm{spin}}G_{\Phi} \), with observable \(A_{\Phi}\) or \(T_{\varphi}\).
\item \textbf{III-Seam:} Microwave dissipation at bonded interfaces, with \( \rho_j=r_{\mathrm{seam}}=g_{\mathrm{seam}}^{-1} \) and geometry functional \( Y_{\mathrm{seam}} \); the leading relation is \( \Pi_{\mathrm{seam}}=r_{\mathrm{seam}}Y_{\mathrm{seam}}=Q^{-1}_{\mathrm{seam}} \), with observable \(Q^{-1}\) or \(T_1\).
\item \textbf{IV-QP:} Quasiparticle relaxation in split form, with environmental variable \( \bar{n}_{qp} \) and trap-geometry factor \( G_{qp}^{(\mathrm{trap})} \); the leading relation is \( \Gamma_{qp}\approx C_{qp}\bar{n}_{qp}G_{qp}^{(\mathrm{trap})} \), with observable \(T_1^{-1}\) and parity-related metrics.
\item \textbf{V-Phonon:} Substrate-mediated acoustic/phonon loss, with effective acoustic state variable \( Z_{ph} \) and geometry functional \( G_{ph} \); the hypothesis-level relation is \( \Pi_{ph}=Z_{ph}G_{ph} \), with observable \(Q^{-1}\) or \(T_1\).
\end{itemize}

\noindent
The seam channel uses the resistivity-like variable $r_{\mathrm{seam}}=g_{\mathrm{seam}}^{-1}$
so that the channel remains in multiplicative form $Q^{-1}=\rho_j G_j$; see
Sec.~\ref{sec:seam}. If desired, $r_{\mathrm{seam}} \equiv g_{\mathrm{seam}}^{-1}$
may be defined explicitly to maintain consistency with literature conventions that
use conductance form. The quasiparticle prescriptor is naturally split into
environmental IV(b) and trap-geometry IV(a) sub-prescriptors; see
Sec.~\ref{sec:qp}. Dimensional closure is discussed in
Appendix~\ref{app:dimensional}.

%% ============================================================
\section{Prescriptor Classes}
\label{sec:channels}

Prescriptors I–III constitute test-ready classes; IV–V define extensions whose validation requires additional environmental control. Five prescriptor classes are defined in this section, organized by their current experimental maturity:

\begin{itemize}
\item \textbf{Established and test-ready:} Prescriptors I, II, and III have complete derivations, dimensional closure, and fully defined experimental protocols.
\item \textbf{Architecturally complete but emerging:} Prescriptor IV requires concurrent environmental monitoring that is  becoming available.
\item \textbf{Forward hypothesis:} Prescriptor V defines a falsifiability test for emerging 3D and flip-chip architectures.
\end{itemize}

These five channels represent a deliberate present-day basis, satisfying three selection criteria: an established experimental signature, a recognized microscopic origin, and a separable coupling structure.

\subsection{Prescriptor I: Curvature-Conditioned Dielectric Loss}
\label{sec:tls}

The microscopic origin of dielectric loss in superconducting circuits was established by Simmonds et al.\ and Cooper et al.\ through spectroscopic resolution of individual TLS-qubit avoided crossings~\cite{simmonds2004,cooper2004}. Those experiments confirmed that atomic-scale interface defects, not bulk or geometric imperfections,set the dominant loss floor, and they motivate the
distribution-resolved statistics central to Prescriptor~I.

Standard participation-ratio models assign a single spatially uniform loss tangent $\tan \delta_i$ to each interface region~\cite{wenner2011,wang2015,gao2008,woods2019}. That assignment is empirically incomplete: process interventions localized to high-field electrode edges produce quality-factor improvements disproportionate to the treated area fraction~\cite{altoe2022,verjauw2021}, which is direct evidence that TLS-mediated loss is spatially heterogeneous and concentrated near structurally severe sites~\cite{muller2019,lisenfeld2019,oconnell2008,simmonds2004,cooper2004}. The appropriate generalization of the participation-ratio formula is
\begin{equation}
Q^{-1} = \frac{\int \epsilon(\mathbf{r}) |\mathbf{E}(\mathbf{r})|^2 \tan \delta_{\text{eff}}(\mathbf{r}) \, dV}{\int \epsilon(\mathbf{r}) |\mathbf{E}(\mathbf{r})|^2 \, dV},
\label{eq:localtan}
\end{equation}
where $\tan \delta_{\text{eff}}(\mathbf{r})$ is now a spatially resolved effective loss tangent.

The present hypothesis is that local electrode curvature serves as a useful reduced descriptor for the broader local state of oxide disorder, strain, and surface under coordination that governs TLS-hosting chemistry~\cite{anderson1972,phillips1972,muller2019}. Curvature does not directly set $\tan \delta_{\text{eff}}$; rather, it correlates with the physical conditions (elastic strain, oxide stoichiometry, tunneling-barrier distribution) that determine TLS density at each site. We write
\begin{equation}
\tan \delta_{\text{eff}}(\kappa) = \tan \delta_0 f(\kappa; \{\alpha_r\}),
\label{eq:curvaturelaw}
\end{equation}
where $\kappa$ is local curvature and $\{\alpha_r\}$ collectively denotes unresolved local chemical state variables.

When the reduced-order approximation is invoked, the natural candidate is the second curvature moment:
\begin{equation}
\mu_2 \equiv \langle R_c^{-2} \rangle = \frac{1}{L} \int_0^L \kappa(s)^2 \, ds,
\label{eq:mu2}
\end{equation}
where $L$ is the relevant edge perimeter and $s$ is arc length. The scalar prescriptor is then
\begin{equation}
\Pi_{\text{I}} = \mu_2 G_{\text{I}},
\label{eq:prescriptorI}
\end{equation}
where $G_{\text{I}}$ is the geometry-derived edge-field weighting functional obtained from finite-element electrostatic solutions~\cite{wenner2011,wang2015,woods2019}.

Using $\mu_2$ as the dominant reduced curvature prescriptor is a direct, falsifiable hypothesis. Three concurring physical mechanisms suggest it is the optimal candidate. First, elastic strain amplification: strain energy density at a geometric cusp scales as $R_c^{-2}$, directly modifying TLS asymmetry energies and tunneling barriers~\cite{anderson1972,phillips1972,lisenfeld2019}. Second, oxide disorder accumulation: high-curvature sites preferentially accumulate chemically disordered native oxide due to surface-energy anisotropy and undercoordination~\cite{verjauw2021,murthy2022,bal2024}. Third, field-disorder overlap: the local energy density $|\mathbf{E}|^2$ is enhanced at sharp edges so that the overlap between high-field and high-TLS-density regions is multiplicative rather than additive~\cite{wenner2011,gao2008,ganjam2024}. Because all three mechanisms scale with $R_c^{-2}$, the second moment $\mu_2$ is a physically motivated candidate to capture the leading contribution, particularly when the high-curvature tail dominates the decoherence budget. $\mu_2$ is not unique as a curvature descriptor, but it is the lowest-order moment consistent with all three contributing mechanisms simultaneously; higher moments ($\mu_3$, $\mu_4$) provide refinements but are not required at leading order. 

The choice of $\mu_2$ should be interpreted as a reduced-order closure, not as an assertion that curvature alone fully parameterizes TLS chemistry. The intent is narrower: among experimentally accessible distribution-resolved statistics, $\mu_2$ is proposed as the lowest-order moment expected to retain sensitivity to rare, high-severity edge configurations while remaining tractable and transferable across device geometries. Its adequacy is therefore an experimental question to be decided by the pre-committed discrimination protocol.

We note explicitly that FEM solvers already capture electrostatic field enhancement at sharp edges through the geometry of $G_{\text{I}}$; introducing an additional curvature weight on $|\mathbf{E}|^2$ would double-count this enhancement. Curvature enters correctly through its influence on the local chemistry in $\tan \delta_{\text{eff}}$, not through an additional field-concentration factor. Two candidate functional forms for $f$ discriminate between mechanisms. The linear tail-weight model is $\tan \delta_{\text{eff}} = \tan \delta_0 (1 + \alpha \mu_2)$ with $\alpha$ in units of m$^2$. The exponential hotspot model is $\tan \delta_{\text{eff}} = \tan \delta_0 \exp(\beta \mu_2)$ with $\beta$ in units of m$^2$; both recover $\tan \delta_0$ as $\mu_2 \to 0$. Discrimination between these models requires a process-split series with at least four distinct values of $\mu_2$ spanning a factor of $\sim 3$ in the statistic.

The microstructural state variable $\rho_{\text{I}} = \mu_2$ is extractable from FIB-SEM curvature histograms (lateral resolution $2$-$5$\,nm) supplemented by STEM for the high-curvature tail~\cite{murthy2022}. In the present work, $\kappa(s)$ is used operationally as a cross-sectional edge-curvature proxy sampled on perimeter-normal slices and assembled into an arclength-weighted statistic over the electrode edge, rather than as a full 3D reconstruction of principal curvatures. Because the relevant curvature distribution may be heavy-tailed, sampling density should be set by convergence of $\mu_2$ and the upper-tail quantiles, not by nominal perimeter coverage alone. For a transmon electrode perimeter of order 1~mm, sampling on the order of 100-200 cross-sectional sites is estimated to yield $\mu_2$ with statistical uncertainty below approximately 10\%; depending on the spatial distribution of curvature heterogeneity along the electrode perimeter; the required sample count scales with the variance of the local curvature distribution and should be verified empirically for each process split. The geometry functional $G_{\text{I}}$ is obtained from a standard FEM electrostatic solve~\cite{wenner2011,wang2015,woods2019}.

\begin{figure*}[t]
\centering
\includegraphics[width=\textwidth]{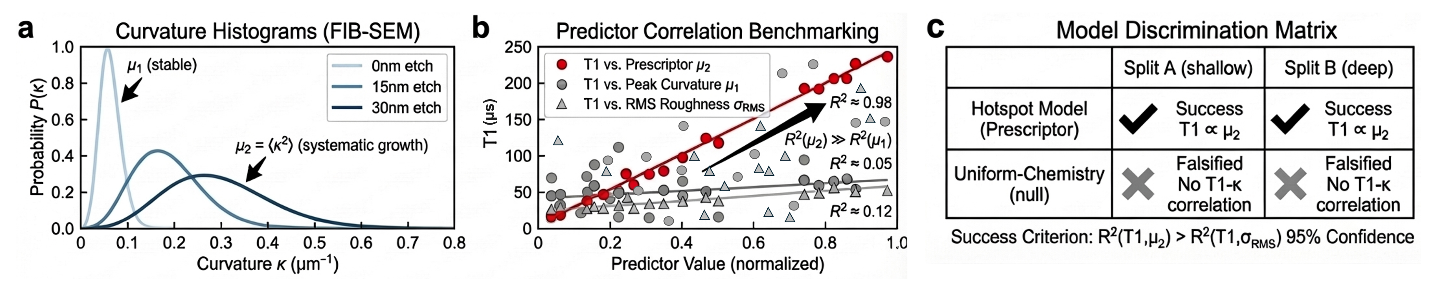}
\caption{\textbf{Prescriptor I extraction pipeline (representative extraction workflow and expected scaling behavior):} format for Part II validation. (a) FIB-SEM curvature histograms for a process-split series with trench-edge etch depth varied $0$-$30$\,nm; the series produces $\mu_2$ values spanning a controlled range at fixed device geometry. (b) Schematic correlation of $T_1$ with three candidate microstructural predictors: $\mu_2$ (second curvature moment), $\mu_1$ (first curvature moment), and $R_{\text{RMS}}$ (RMS roughness). (c) Pre-committed model-discrimination matrix. Success criterion: $R^2(T_1, \mu_2) > R^2(T_1, R_{\text{RMS}})$ at 95\% confidence across the full etch-depth split series. Failure of this criterion would indicate that curvature is not a sufficient reduced descriptor in this processing regime.Expected trends are based on reported process-induced curvature variations in superconducting devices.}
\label{fig:curvature}
\end{figure*}

As illustrated in FIG.~\ref{fig:curvature}, this framework allows for independent evaluation of different spatial moments across a controlled geometric split series, providing a predictive protocol for future device design.

\noindent\textit{Falsification criterion:} $R^2(T_1,\mu_2) \leq R^2(T_1, R_\mathrm{RMS})$ across the full etch-depth split series, or systematic failure of the $\rho$-ratio and $G$-ratio checks of Sec.~\ref{sec:falsifiability}, would indicate that curvature is not a sufficient reduced descriptor in this processing regime.

Prescriptor I is thus a canonical test case that separates a measurable microstructural statistic ($\mu_2$) from a computable geometric functional ($G_\text{I}$), allowing a direct test of predictive transfer across process and geometry splits.

\subsection{Prescriptor II: Surface-Spin Flux Noise}
\label{sec:spin}

Flux noise in superconducting qubits exhibits an approximately $1/f$ spectral density over broad frequency ranges~\cite{koch2007,sendelbach2008,yoshihara2006}.  Fluctuating surface or near-surface magnetic moments remain among the most extensively supported microscopic models for this behavior~\cite{martinis2005,bilmes2020,sendelbach2008}.

\noindent\textit{What is measured ($\rho_j$):} surface or near-surface spin density $\rho_\mathrm{spin}$~[m$^{-2}$], extracted from witness-sample EPR measurements independently of device geometry. \textit{What is computed ($G_j$):} magnetostatic coupling integral $G_\Phi$~[T$^2$A$^{-2}$m$^2$] from a single FEM solve of the SQUID loop field. \textit{Leading observable:} $A_\Phi$ or $T_\varphi$.  \textit{Falsification criterion:} $A_\Phi$ fails to scale as $\rho_\mathrm{spin} G_\Phi$ across the $2\!\times\!2$ protocol; or the row-ratio and column-ratio checks are mutually inconsistent. We adopt the angular-frequency spectral convention.

\begin{equation}
  S_\Phi(\omega) = \frac{A_\Phi}{|\omega|},
  \label{eq:noisespectrum}
\end{equation}

where $A_\Phi$ carries units of~$\Phi_0^2$, and note that the precise numerical value of $A_\Phi$ depends on whether a one-sided or two-sided spectral convention is used; this choice must be stated explicitly in any quantitative comparison.

The prescriptor architecture separates the question of \textit{how many} spins are present from the question of \textit{how strongly} they couple to the SQUID loop.  By the electromagnetic reciprocity principle, a unit current $I$ flowing in the SQUID loop produces a magnetic field $\mathbf{B}_\mathrm{circ}(\mathbf{r})$ at surface point $\mathbf{r}$.  The flux coupling of a single spin at position $\mathbf{r}$ to the loop is $\phi(\mathbf{r}) = \mathbf{B}_\mathrm{circ}(\mathbf{r})/I$ [units: T$\cdot$A$^{-1}$].  After orientation averaging and under the assumption that spin fluctuations are spatially uncorrelated, the geometry functional is
\begin{equation}
  G_\Phi
  = \int_{\partial\Omega_\mathrm{spin}}
    \left|\frac{\mathbf{B}_\mathrm{circ}(\mathbf{r})}{I}\right|^2
    dA
  \quad [\text{T}^2\cdot\text{A}^{-2}\cdot\text{m}^2],
  \label{eq:GPhi}
\end{equation}
computable from a single magnetostatic FEM solve.  The spin prescriptor is then
\begin{equation}
  \Pi_\mathrm{spin} = \rho_\mathrm{spin}\,G_\Phi,
  \qquad
  A_\Phi \approx C_\Phi\,\rho_\mathrm{spin}\,G_\Phi,
  \label{eq:prescriptorII}
\end{equation}
where $C_\Phi = \mu_B^2 / \Phi_0^2$ absorbs Bohr-magneton factors, orientation-averaging coefficients, and the chosen spectral convention prefactor.

A dimensional consistency check is instructive. $\rho_\mathrm{spin}$\,[m$^{-2}$] $\times$
$G_\Phi$\,[T$^2$A$^{-2}$m$^2$] $\times$ $\mu_B^2$\,[J$^2$T$^{-2}$] $= $ J$^2$A$^{-2}$ = Wb$^2 = \Phi_0^2$, confirming closure.

\subsection{Prescriptor III: Seam and Interface Dissipation}
\label{sec:seam}

We define a seam as any bonded metal-metal or metal-dielectric interface that lies in the microwave current path~\cite{romanenko2020,brecht2016,reagor2016}. What is measured: $j$: seam resistivity-like parameter $r_{\text{seam}}$ [$\Omega$m], extracted from independent witness structures (e.g., CPW resonators) fabricated with the identical bonding process. What is computed: $G_j$: seam current participation factor $Y_{\text{seam}}$ [Sm$^{-1}$], derived from FEM surface current distributions along the seam contour. Leading observable: $Q^{-1}_{\text{seam}}$ or $T_1$. Falsification criterion: $Q^{-1}$ scaling fails to match the product $r_{\text{seam}} Y_{\text{seam}}$ across a process-geometry split, indicating uncaptured interface effects, non-linear contact resistance, or breakdown of the spatially uniform seam assumption.

For a resonator or qubit operating at angular frequency $\omega$ with stored energy $U_{\text{stored}}$, we define $g_{\text{seam}}$ as the seam conductance per unit length [Sm$^{-1}$]. With $J_s(s)$ the microwave surface current density along the seam contour [Am$^{-1}$], the power dissipated at the seam is:
\begin{equation}
P_{\text{loss}} = \frac{1}{2} \int \frac{|J_s(s)|^2}{g_{\text{seam}}} \, ds.
\label{eq:Ploss}
\end{equation}
The quality-factor contribution is $Q^{-1}_{\text{seam}} = P_{\text{loss}} / (\omega U_{\text{stored}})$, from which we define the corresponding seam participation factor:
\begin{equation}
Y_{\text{seam}} = \frac{1}{2\omega U_{\text{stored}}} \int |J_s(s)|^2 \, ds \quad [\text{Sm}^{-1}].
\label{eq:Yseam}
\end{equation}

To preserve the multiplicative separable form $Q^{-1}_{\text{seam}} = \rho_j G_j$, we define the material variable as the seam resistivity-like quantity $r_{\text{seam}} \equiv g_{\text{seam}}^{-1}$ [$\Omega$m], giving:
\begin{equation}
\Pi_{\text{seam}} = r_{\text{seam}} \, Y_{\text{seam}} = Q^{-1}_{\text{seam}},
\label{eq:prescriptorIII}
\end{equation}
which is dimensionless as required.

\subsection{Prescriptor IV: Quasiparticle Relaxation (Split)}
\label{sec:qp}

Non-equilibrium quasiparticles are a primary contributor to $T_1$ loss~\cite{catelani2011,pop2014,gordon2022,serniak2018}, but the prescriptor must be split into two physically distinct sub-components to avoid conflating variables that are controlled by entirely different interventions.

\noindent\textit{What is measured ($\rho_j$):} non-equilibrium quasiparticle density $\bar{n}_{qp}$ (environmental sub-prescriptor), inferred from continuous parity-switching or charge-dispersion measurements. \par\noindent\textit{What is computed ($G_j$):} trap-geometry factor $G^{(\mathrm{trap})}_{qp}$ (geometry sub-prescriptor), reflecting quasiparticle diffusion and capture cross-sections. \par\noindent\textit{Leading observable:} $T_1^{-1}$ parity-correlated fluctuations. \par\noindent\textit{Falsification criterion:} Parity-switching rate (confirming $\bar{n}_{qp}$ is stable) remains constant while varying $G^{(\mathrm{trap})}_{qp}$, but the measured $T_1^{-1}$ fails to scale proportionally.

\textbf{Prescriptor IV(a): trap-geometry sub-prescriptor.} The rate at which a quasiparticle tunnels through the Josephson junction and is captured by a normal-metal trap depends on the spatial overlap of the quasiparticle wavefunction with the junction region, the trap geometry, and the associated diffusion pathways. We denote this factor $G_{qp}^\mathrm{(trap)}$. The split $n_{\text{qp}}$ - $G_{\text{qp}}^{\text{trap}}$ architecture is consistent with diffusion-and-trapping models~\cite{riwar2016normal} in which evacuation time depends on trap size, diffusion constant, and device geometry, and saturates once trap size exceeds a geometry-dependent scale. In this language, separability is expected to fail when depletion regions overlap, backflow becomes appreciable, or burst-driven phonon dynamics transiently change the effective quasiparticle environment during the measurement window.

\textbf{Prescriptor IV(b): environmental sub-prescriptor.} The non-equilibrium quasiparticle density $\bar{n}_{qp}$ is set by the balance of pair-breaking from infrared photon loading, cosmic-ray events, and stray radiation~\cite{vespalainen2020,mcewen2022}. This is an environmental state variable not under direct fabrication control. The leading-order effective factorization is
\begin{equation}
  \Gamma_{qp} \approx C_{qp}\,\bar{n}_{qp}\,G_{qp}^\mathrm{(trap)},
  \label{eq:prescriptorIV}
\end{equation}
under quasistatic operating conditions. Validation of IV(a) requires
parity-switching or charge-dispersion measurements to verify that
$\bar{n}_{qp}$ remains constant while $G_{qp}^\mathrm{(trap)}$ is
varied.

\subsection{Prescriptor V: Phonon Reservoir Topology (Hypothesis)}
\label{sec:phonon}

As flip-chip and multi-chip superconducting modules become standard, substrate-mediated acoustic phonon loss is anticipated to become a measurable coherence channel~\cite{brecht2016,gordon2022}. 
\noindent\textit{What is measured ($\rho_j$):} effective acoustic state variable $Z_{ph}$, capturing boundary impedance or specific substrate defect characteristics. \par\noindent\textit{What is computed ($G_j$):} dimensionless electromagnetic-to-acoustic overlap functional $G_{ph}$. \par\noindent\textit{Leading observable:} $Q^{-1}$ or $T_1$. \par\noindent\textit{Falsification criterion:} $T_1$ fails to vary with substrate thickness or mechanical boundary condition at fixed electromagnetic layout, or the variation is inconsistent with the $\Pi_{ph} = Z_{ph} G_{ph}$ ratio prediction. Recent observations of stress-induced phonon bursts linked to quasiparticle poisoning~\cite{anthony2024stress} also support treating the phonon channel as a hypothesis-level state-variable times coupling-functional factorization, while underscoring that acoustic boundary conditions and stress relaxation can become experimentally relevant control variables.

We define a fifth channel as a forward hypothesis:
\begin{equation}
  \Pi_{ph} = Z_{ph}\,G_{ph},
  \label{eq:prescriptorV}
\end{equation}
where $Z_{ph}$ is an effective acoustic state variable parameterizing boundary impedance and $G_{ph}$ is a dimensionless electromagnetic-to-acoustic overlap functional. The falsifiable prediction is that $T_1$ will vary systematically with substrate thickness or mechanical boundary condition at fixed electromagnetic layout. This channel is included to define a forward experimental test rather than a confirmed contribution.

%% ============================================================
\section{Cross-Channel Coupling and Observational Classification}
\label{sec:coupling}

The standard operational decomposition
\begin{equation}
  \frac{1}{T_2} = \frac{1}{2T_1} + \frac{1}{T_\varphi}
  \label{eq:T2}
\end{equation}
remains the fundamental experimental bookkeeping identity, but it
should not be over-interpreted as a universal mechanistic
partition~\cite{krantz2019,kjaergaard2020}. For non-Markovian
processes, most notably $1/f$ flux noise, the extracted pure-dephasing time depends strongly on the filter function of the pulse sequence. Table~\ref{tab:coupling} provides an operational channel classification for the transmon regime.

\begin{table}[t]
\caption{Operational cross-channel coupling map for transmon-class devices.
$\bullet$\,=\,primary coupling; $\circ$\,=\,secondary or indirect coupling.
Assignments are operational in the transmon regime.}
\label{tab:coupling}
\begin{threeparttable}
\footnotesize
\setlength{\tabcolsep}{3pt}
\begin{tabular*}{\columnwidth}{@{\extracolsep{\fill}} l c c c l}
\toprule
Prescriptor & $T_1^{-1}$ & $T_\varphi^{-1}$ & Markovian? & Main caveat \\
\midrule
I - TLS    & $\bullet$ & $\circ$ & Often   & Slow dielectric fluctuations. \\
II - Spin  & $\circ$   & $\bullet$ & No     & Echo vs.\ Ramsey differ. \\
III - Seam & $\bullet$ & $\circ$ & Often   & Primarily relaxation. \\
IV - QP    & $\bullet$ & $\circ$ & Approx. & Parity dynamics. \\
V - Phonon & $\bullet$ & $\circ$ & Unknown & Hypothesis-level. \\
\bottomrule
\end{tabular*}
\begin{tablenotes}[flushleft]
\item Eq.~(\ref{eq:T2}) is used here as a measurement decomposition.
\item Channel assignment is operational rather than universal.
\end{tablenotes}
\end{threeparttable}
\end{table}

%% ============================================================
\section{Falsifiability Protocol: The $2\times2$ Decoupling Matrix}
\label{sec:falsifiability}

\begin{figure*}[t]
\centering
\includegraphics[width=0.96\textwidth]{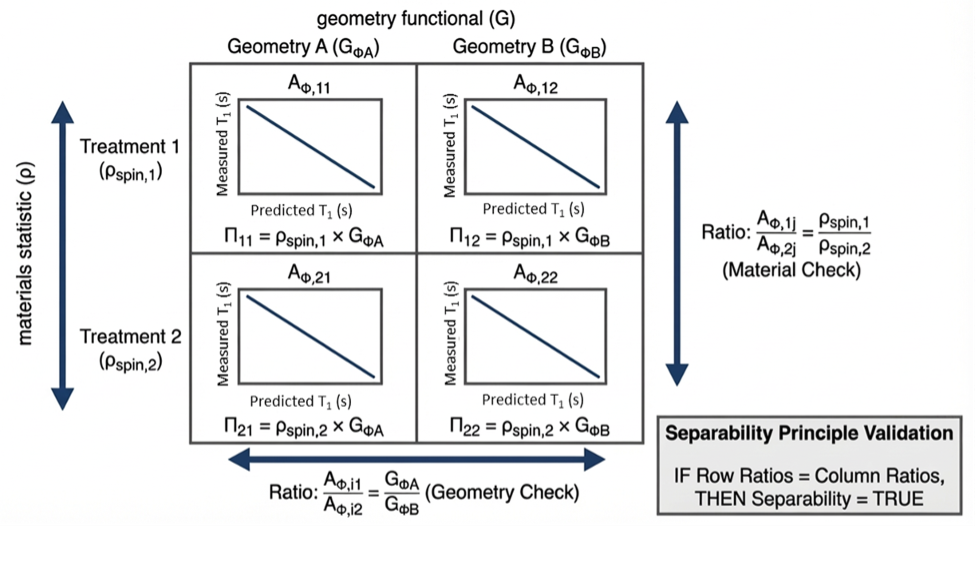}
\caption{
\textbf{Schematic of the \(2\times2\) separability protocol.} Rows correspond to materials treatments that vary the channel-specific microstructural state variable \(\rho_j\) while nominally holding geometry fixed; columns correspond to device geometries that vary the coupling functional \(G_j\) while nominally holding the materials treatment fixed. Each cell defines a pre-committed prescriptor \(\Pi_{mn}=\rho_{j,m}G_{j,n}\) and an independently measured observable \(O^{\mathrm{meas}}_{mn}\). The row-ratio check tests the materials axis and the column-ratio check tests the geometry axis. Agreement of both checks within propagated uncertainty supports the separable reduced-order model in the tested operating regime.
}
\label{fig:2x2protocol}
\end{figure*}

The $2!\times!2$ protocol is the central operational contribution of this work. The $2\!\times\!2$ control design specified here is the minimum experimental structure under which a channel-wise prescriptor can be confirmed or falsified. The microstructural state variable $\rho_j$ and the coupling functional $G_j$ must be varied independently, and the prescriptor predictions committed from independent metrology and field calculations before coherence data are inspected. Without this design, mechanistic claims remain structurally underdetermined. The protocol structure is summarized in FIG.~\ref{fig:2x2protocol}.

Let two materials treatments or process conditions (rows) produce witness-sample estimates $\rho_{j,a}$ and $\rho_{j,b}$, and let two device geometries (columns) produce FEM estimates $G_{j,A}$ and $G_{j,B}$. The separable prescription yields four pre-committed predicted observables:
\begin{equation}
\begin{aligned}
\mathcal{O}_{aA}^\mathrm{pred} &= C_j\,\rho_{j,a}\,G_{j,A},\\
\mathcal{O}_{aB}^\mathrm{pred} &= C_j\,\rho_{j,a}\,G_{j,B},\\
\mathcal{O}_{bA}^\mathrm{pred} &= C_j\,\rho_{j,b}\,G_{j,A},\\
\mathcal{O}_{bB}^\mathrm{pred} &= C_j\,\rho_{j,b}\,G_{j,B}.
\end{aligned}
\label{eq:predictions}
\end{equation}
After measurements yield $\mathcal{O}_{mn}^\mathrm{meas}$, separability is assessed through two ratio checks:
\begin{align}
  \text{Row-ratio test:}\quad
  &\frac{\mathcal{O}_{1A}^\mathrm{meas}}
        {\mathcal{O}_{2A}^\mathrm{meas}}
  \approx
  \frac{\mathcal{O}_{1B}^\mathrm{meas}}
        {\mathcal{O}_{2B}^\mathrm{meas}}
  \approx
  \frac{\rho_{j,1}}{\rho_{j,2}},
  \label{eq:rowcheck} \\
  \text{Column-ratio test:}\quad
  &\frac{\mathcal{O}_{1A}^\mathrm{meas}}
        {\mathcal{O}_{1B}^\mathrm{meas}}
  \approx
  \frac{\mathcal{O}_{2A}^\mathrm{meas}}
        {\mathcal{O}_{2B}^\mathrm{meas}}
  \approx
  \frac{G_{j,A}}{G_{j,B}}.
  \label{eq:columncheck}
\end{align}

Unlike conventional fitting approaches, this protocol requires that all ratios be satisfied simultaneously without adjustable parameters. Failure of either ratio condition provides direct evidence of missing variables or breakdown of separability. Agreement supports the separable reduced-order model in the tested operating regime; systematic deviation indicates cross-perturbation, model incompleteness, or breakdown of the dilute-defect approximation.

\subsection{Operational decision rule}

Before coherence data acquisition, the experimenter specifies a channel-dependent tolerance bound~$\epsilon_j$, a statistical confidence level, and the explicit uncertainty model used to combine errors from metrology, field simulation, and device-to-device scatter. The reduced-order model is declared \textit{supported} only if: (1)~both the row residual and column residual are statistically compatible with zero, meaning the fractional deviation of each measured ratio from its predicted value falls within the pre-registered tolerance $\epsilon_j$ at the specified confidence level (e.g., 95\% CI from propagated metrology, FEM numerical uncertainty, and device-to-device scatter combined in quadrature); and (2)~the measured ratios are compatible with the independently predicted $\rho_j$ and $G_j$ ratios within those pre-registered bounds. Both conditions must be satisfied simultaneously; passing one axis while failing the other constitutes a falsification of separability for that channel.

%% ============================================================
\section{Minimum-Dataset Specification and Literature Alignment}
\label{sec:mds}

For the prescriptor framework to support cumulative inference
across laboratories, datasets must be comparable through a common
reporting standard.  We define a Minimum-Dataset Specification that
distinguishes datasets capable of supporting a quantitative
prescriptor test from those informative only for qualitative
trend identification.

The specification requires reporting in three categories:
\begin{enumerate}
\item \textbf{Microstructural State Variables ($\rho_j$):}\\
  Independent measurement of the relevant structural statistic
  (e.g., $\mu_2$ via FIB-SEM, $\rho_\mathrm{spin}$ via EPR,
  $r_\mathrm{seam}$ via CPW) with specified statistical bounds,
  acquired from witness samples processed identically to the
  device wafer.
\item \textbf{Geometry Coupling Functionals ($G_j$):}\\
  Explicit specification of the mode-volume definitions, boundary
  conditions, and FEM procedures used to compute $G_I$, $G_\Phi$,
  or $Y_\mathrm{seam}$, allowing third-party reproduction of the
  coupling factors.
\item \textbf{Coherence Observables ($\mathcal{O}_j$):}\\
  Device-level observables reported with their associated protocol
  context (e.g., stating whether $T_\varphi$ was acquired via
  Ramsey or echo, specifying the measurement window for $T_1$
  fluctuations, and confirming parity stability for QP tests).
\end{enumerate}

%% ============================================================
\section{Validation Roadmap}
\label{sec:roadmap}

The present manuscript establishes the formal architecture of the prescriptor framework.  Coordinated experimental validation will be reported in Part~II, executing the $2\!\times\!2$ protocol across four principal axes:
\begin{enumerate}
\item \textbf{Prescriptor I (TLS/Curvature):} Independent variation of electrode edge curvature ($\mu_2$) via controlled etch chemistry against geometric participation ratio ($G_I$) via gap-width scaling.
\item \textbf{Prescriptor II (Spin/Flux):} Independent variation of surface spin density ($\rho_\mathrm{spin}$) via ambient oxidation time against magnetic coupling ($G_\Phi$) via SQUID-loop area modulation.
\item \textbf{Prescriptor III (Seam):} Independent variation of interface processing ($r_\mathrm{seam}$) against seam current participation ($Y_\mathrm{seam}$) via geometry modulation.
\item \textbf{Prescriptor IV (QP):} Controlled variation of trap geometry ($G_{qp}^\mathrm{(trap)}$) while maintaining continuous monitoring of the parity-switching rate to verify environmental stationarity ($\bar{n}_{qp}$).
\end{enumerate}

%% ============================================================
\section{Discussion and Conclusion}
\label{sec:discussion}

The transition from empirical observation to predictive engineering in superconducting quantum circuits motivates a methodological shift: improvements in coherence time must be decomposed into independently measurable microstructural statistics and independently computable geometry functionals.  Without that separation, process optimization remains fundamentally underdetermined because chemistry, topology, and geometry are typically perturbed together.

For theorists, the intended contribution is not a replacement of microscopic decoherence models, but a disciplined mapping from those models to independently accessible materials variables and geometry functionals suitable for cross-wafer and cross-geometry inference.

The core framework - \textit{separability and falsifiability}, summarized by the prescriptor $\Pi_j = \rho_j G_j$ - provides the required structure; the five prescriptor classes are the first candidate instantiations. It identifies the leading-order conditions under which the defect field can be coarse-grained into a useful scalar statistic. It operationalizes falsifiability through a pre-committed $2\!\times\!2$ decoupling matrix, ensuring that the model is tested prospectively rather than fitted retroactively. It establishes a Minimum-Dataset Specification to guide future reporting.

A recurring pattern in successful structure-property formulations is the recognition that macroscopic behavior factorizes into
independently measurable and independently computable variables; the lasting contribution is not a new instrument but the disciplined identification of which two variables to separate. By isolating the microstructural state variables $\rho_j$ from the device-specific coupling functionals $G_j$, the prescriptor architecture allows materials scientists to characterize and optimize quantum-relevant disorder distributions on witness samples and to test whether those improvements transfer predictively across circuit geometries for which the corresponding coupling functional can be computed within the separability regime established here.

The prescriptor classes developed here are derived for the transmon regime ($E_J/E_C \gg 1$); their specific coupling kernels do not transfer directly to other architectures without re-derivation. The separable $\rho \!\cdot\! G$ structure, however, may extend more broadly. In fluxonium, the superinductance loop geometry modifies $G_{\Phi}$, while $\rho_{\mathrm{spin}}$ on the wire surface remains independently measurable. In color-center devices, surface-spin noise couples through the emitter's position relative to the surface, providing a close analogue of Prescriptor II. In semiconductor spin qubits, interface-defect-mediated charge noise couples through an electrostatic potential set by gate geometry, providing a close analogue of Prescriptor I. These are structural analogies, not validated separability claims; establishing $\rho \!\cdot\! G$ factorization in any of these systems requires a platform-specific falsifiability program. The immediate test of this framework lies not in its formulation, but in whether its predictions survive independent process and geometry variation.

\begin{acknowledgments}
The authors thank A.\ Romanenko and A.\ Grassellino for stimulating discussions, and colleagues at the SQMS Center and the Northwestern NU\textit{ANCE} Center. We appreciate the contribution by Mr.\ Amil Dravid, Ph.D.\ student at BAIR/UC Berkeley, for timely review and constructive input to this work. This work was supported by the U.S.\ Department of Energy, Office of Science, National Quantum Information Science Research Centers, Superconducting Quantum Materials and Systems Center (SQMS), under Contract No.\ 89243024CSC000002. Fermilab is operated by Fermi Forward Discovery Group, LLC under Contract No.\ 89243024CSC000002 with the U.S.\ Department of Energy, Office of Science, Office of High Energy Physics. This work made use of the NUANCE Center at Northwestern University, which has received support from the SHyNE Resource (NSF ECCS-2025633), the IIN, and Northwestern's MRSEC program (NSF DMR-2308691).
\end{acknowledgments}

\begin{aidisclosure}
AI-assisted tools were used for language and organizational refinement during manuscript preparation. Scientific content, conceptual development, mathematical analysis, and editorial judgment belong to the authors.
\end{aidisclosure}

%% ============================================================
\bibliography{references}

% \clearpage
\appendix
\section{Dimensional Closure of Prescriptors}
\label{app:dimensional}

The physical consistency of the prescriptor framework requires that the product $\Pi_j = \rho_j G_j$ map dimensionally to the device level observable $\mathcal{O}_j$ through a prefactor $C_j$ that contains only fundamental constants and protocol-specific dimensionless numbers.  All prescriptor forms satisfy dimensional closure, as verified below.

\textbf{Prescriptor I (TLS).} Observable: $Q^{-1}$ (dimensionless). State variable: $\rho_{\text{I}} = \mu_2 = \langle R_c^{-2}\rangle$, units m$^{-2}$. Geometry functional: $G_I$ scales as a participation length $\ell \sim$ m$^2$. Product: $\mu_2 \cdot G_I$ [m$^{-2} \cdot$ m$^2$] = dimensionless.

\textbf{Prescriptor II (Spin).} Observable: $A_\Phi$ in $S_\Phi(\omega) = A_\Phi/|\omega|$, units $\Phi_0^2$ (one-sided convention; see Sec.~\ref{sec:spin}).
\begin{align*}
\rho_\mathrm{spin} &\quad [\mathrm{m}^{-2}] \\
G_\Phi = \int |\mathbf{B}_\mathrm{circ}/I|^2\,dA
  &\quad [\mathrm{T}^2\cdot\mathrm{A}^{-2}\cdot\mathrm{m}^2] \\
C_\Phi = \mu_B^2/\Phi_0^2
  &\quad [\mathrm{J}^2\cdot\mathrm{T}^{-2}\cdot\mathrm{Wb}^{-2}]
\end{align*} 
Product step-by-step:
$[\mathrm{m}^{-2}]\cdot[\mathrm{T}^2\mathrm{A}^{-2}\mathrm{m}^2]
= \mathrm{T}^2\mathrm{A}^{-2}$;
then $\times\,[\mathrm{J}^2\mathrm{T}^{-2}\mathrm{Wb}^{-2}]
= \mathrm{J}^2\mathrm{A}^{-2}\mathrm{Wb}^{-2}
= \mathrm{Wb}^2 \cdot \mathrm{Wb}^{-2}
\cdot \Phi_0^2 = \Phi_0^2$.

\textbf{Prescriptor III (Seam).} Observable: $Q^{-1}_\mathrm{seam}$ (dimensionless). State variable: $r_\mathrm{seam} = g_\mathrm{seam}^{-1}$ [$\Omega\cdot$m]. Geometry functional: $Y_\mathrm{seam}$ [S/m] = [$\Omega^{-1}$m$^{-1}$]. Product: $r_\mathrm{seam} \cdot Y_\mathrm{seam}$ [$\Omega$m]$\cdot$[$\Omega^{-1}$m$^{-1}$] = dimensionless.

\textbf{Prescriptor IV (Quasiparticle).} Observable: $\Gamma_{qp}$ contribution to $T_1^{-1}$, units s$^{-1}$. State variable: $\bar{n}_{qp}$ [m$^{-3}$]. Geometry functional: $G_{qp}^\mathrm{(trap)}$ carries units of capture-rate kinematic factor $C_{qp}G_{qp}^\mathrm{(trap)}$ [m$^3$s$^{-1}$]. Product: $\bar{n}_{qp} \cdot C_{qp}G_{qp}^\mathrm{(trap)}$ [m$^{-3}$][m$^3$s$^{-1}$] = s$^{-1}$.

\textbf{Prescriptor V (Phonon).} Observable: $Q^{-1}$ (dimensionless). State variable: $Z_{ph}$ (dimensionless acoustic reflection coefficient). Geometry functional: $G_{ph}$ (dimensionless EM - acoustic overlap integral). Product: $Z_{ph} \cdot G_{ph}$ = dimensionless. 
\section*{Appendix B: Inverse Design and Coherence Budgeting}

The factorization \(\Pi_j = \rho_j G_j\) can be used in an inverse-design sense. Because \(\rho_j\) and \(G_j\) are independently measurable and computable, a target device performance can be back-calculated into specific materials requirements and geometric constraints.

This inverse logic is valid within the separability regime of (Eq.~(\ref{eq:goldenrule} of the main text) and should be interpreted as a leading-order engineering guide, not as a globally exact optimization. This appendix develops the inverse-design logic, identifies the constraints that bound its applicability, and illustrates the approach with a coherence-budgeting example.

\subsection*{I. The Total Coherence Budget}

We define \(T_{1,\mathrm{target}}\) as the minimum relaxation time required for a specified device-level function, for example the \(T_1\) floor needed for a target logical error rate or for a given gate depth. It is an externally specified engineering constraint, not an emergent quantity of the prescriptor formalism.

Once \(T_{1,\mathrm{target}}\) is fixed, the total allowable loss rate is
\begin{equation}
\Gamma_{\mathrm{total}} = \frac{1}{T_{1,\mathrm{target}}}.
\tag{Eq. B1}
\end{equation}

For a device to satisfy this target, the sum of all independent loss-rate contributions must obey
\begin{equation}
\frac{1}{T_{1,\mathrm{target}}}
\ge
\sum_j \Gamma_j
\approx
\sum_j C_j \bigl(\rho_j \cdot G_j\bigr) + \Delta_{\mathrm{total}},
\tag{Eq. B2}
\end{equation}
where the sum runs over the five prescriptor channels, \(C_j\) are channel-dependent prefactors defined in the Section~\ref{sec:architecture} of the main text, and \(\Delta_{\mathrm{total}}\) collects higher-order residuals and unformalized channels.

Within the separability regime, \(\Delta_{\mathrm{total}}\) should remain subdominant. A persistently large \(\Delta_{\mathrm{total}}\) signals either an unformalized loss mechanism or a breakdown of the factorization itself; it is therefore diagnostic rather than nuisance structure.

The experimenter treats \(T_{1,\mathrm{target}}\) as the independent variable and distributes the loss budget across channels. The budget is zero-sum: relaxing one channel allowance tightens the others, but the total must not exceed \(1/T_{1,\mathrm{target}}\). In this way, coherence engineering becomes a constrained allocation problem rather than open-ended process optimization.

\subsection*{II. Back-Calculation of Microstructural Requirements}

Given a fixed device geometry, often constrained by readout, coupling, or frequency requirements, the geometry functionals \(G_j\) are determined. The inverse problem then becomes: \textit{what microstructural state variables \(\rho_j\) are required to meet the coherence target?}

For each channel, the prescriptor formalism defines an upper bound on \(\rho_j\), namely the maximum allowable value of the materials-side state variable that remains consistent with the allocated loss budget.

For Prescriptor I (curvature-modulated dielectric loss), the required second curvature moment is bounded by
\begin{equation}
\mu_2^{\mathrm{limit}}
\le
\frac{Q_{\mathrm{target}}^{-1} - \sum_{j\neq I} Q_j^{-1}}
{C_I G_I}.
\tag{Eq. B3}
\end{equation}

This gives a practical metrological target. If FIB-SEM curvature histograms of witness samples yield \(\mu_2 > \mu_2^{\mathrm{limit}}\), then the formalism predicts that the coherence target will not be met by this channel alone, regardless of other process improvements. Conversely, if \(\mu_2\) already lies well below the upper bound, further curvature reduction offers diminishing returns and the budget is likely limited elsewhere.

The same logic applies to the remaining channels. For Prescriptor II, the critical spin density \(\rho_{\mathrm{spin}}^{\mathrm{limit}}\) is determined by the flux-noise allowance and the computed \(G_\Phi\). For Prescriptor III, the critical seam resistivity \(r_{\mathrm{seam}}^{\mathrm{limit}}\) is determined by the seam-loss allowance and the computed \(Y_{\mathrm{seam}}\).

In each case, the back-calculated upper bound is a measurable quantity on a witness sample, giving a go/no-go criterion before device fabrication begins. These bounds should be interpreted within the propagated uncertainty of both witness-sample metrology and FEM computation; they define tolerance bands, not sharp thresholds.

\subsection*{III. Illustrative Coherence Budget}

To make the inverse-design logic concrete, consider a transmon qubit with $T_{1,\mathrm{target}} = 1~\mathrm{ms},$ so that the total loss budget is $\Gamma_{\mathrm{total}} = 10^3~\mathrm{s}^{-1}$.

Suppose the device geometry is fixed by readout and coupling requirements, so all \(G_j\) are known from FEM. A representative channel allocation is shown in Table~\ref{tab:B1_budget}.

\setcounter{table}{0}
\renewcommand{\thetable}{B\arabic{table}}
\begin{table*}[t]
\caption{Illustrative coherence budget for $T_1 = 1~\mathrm{ms}$. 
Allocations reflect engineering priors from published loss-channel analyses.}
\label{tab:B1_budget}
\begin{threeparttable}
\footnotesize
\setlength{\tabcolsep}{6pt}
\renewcommand{\arraystretch}{1.3}
\begin{tabular*}{\textwidth}{@{\extracolsep{\fill}} l c c l l l }
\toprule
\textbf{Channel} & \textbf{Loss Allowance} & \textbf{$G_j$ (FEM)} & \textbf{$\rho_j$ upper bound} & \textbf{Measurement} & \textbf{Rationale} \\
\midrule
I - TLS   & $400~\mathrm{s}^{-1}$ (40\%) & known & $\mu_2 \le \mu_2^{\mathrm{lim}}$ 
           & FIB-SEM     & Dominant \\
II - Spin & $200~\mathrm{s}^{-1}$ (20\%) & known & $\rho_{\mathrm{spin}} \le \rho_{\mathrm{spin}}^{\mathrm{lim}}$ 
           & EPR          & Measurable \\
III - Seam & $200~\mathrm{s}^{-1}$ (20\%) & known & $r_{\mathrm{seam}} \le r_{\mathrm{seam}}^{\mathrm{lim}}$ 
            & CPW witness & Arch.-dependent \\
IV - QP   & $100~\mathrm{s}^{-1}$ (10\%) & known & $\bar{n}_{\mathrm{qp}} \le \bar{n}_{\mathrm{qp}}^{\mathrm{lim}}$ 
           & Parity       & Partly mitigatable \\
V - Phonon+$\Delta$ & $100~\mathrm{s}^{-1}$ (10\%) & TBD & (margin)
           & ---          & Residual \\
\bottomrule
\end{tabular*}
\end{threeparttable}
\end{table*}

\begin{table*}[t]
\caption{Qualitative co-optimization conflict matrix. Signs indicate direction of change in coupling functional. Quantitative values require device-specific FEM.}
\label{tab:B2_conflict}
\begin{threeparttable}
\footnotesize
\setlength{\tabcolsep}{6pt}
\renewcommand{\arraystretch}{1.3}
\begin{tabular*}{\textwidth}{@{\extracolsep{\fill}} l c c c l @{}}
\toprule
\textbf{Geometric change} & \textbf{$G_I$ (TLS)} & \textbf{$G_\Phi$ (Spin)} & \textbf{$y_{\mathrm{seam}}$ (Seam)} & \textbf{Net effect} \\
\midrule
Enlarge pad area   & $+$ & $-$ & 0 & Trade-off (I vs II) \\
Increase gap width & $+$ & 0   & 0 & Favorable \\
Enlarge SQUID loop & 0   & $-$ & 0 & Unfavorable for II \\
Reduce seam length & 0   & 0   & $+$ & Favorable \\
Thin substrate     & 0   & 0   & 0/$-$ & V-dependent \\
\bottomrule
\end{tabular*}
\end{threeparttable}
\end{table*}
The allocations in Table~\ref{tab:B1_budget} are not arbitrary, TLS-mediated dielectric loss is consistently the dominant channel in state-of-the-art transmons ~\cite{cooper2004,wang2015,anthony2024stress}, justifying $\approx$40\%. Flux noise and seam dissipation each contribute measurably ~\cite{koch2007,sendelbach2008,yoshihara2006}, warranting $\approx$20\% each. Quasiparticle poisoning is partially mitigatable through shielding ($\approx$10\%). The remainder ($\approx$10\%) is margin for phonon loss and unformalized residuals. These are starting-point allocations; the point is the budgeting structure, not the specific numbers.

Once the budget is allocated and the geometry fixed, each channel defines a pre-fabrication criterion that can be evaluated on witness samples before a single qubit is built.

\subsection*{IV. Geometric Sensitivity and Design-Parameter Prioritization}

When \(\rho_j\) is constrained by materials availability or process maturity, the remaining design freedom is geometric. The efficiency of a geometric intervention is governed by
\begin{equation}
\mathcal{S}_{G,j} = \frac{\partial G_j}{\partial p},
\tag{Eq. B4}
\end{equation}
where \(p\) is a vector of geometric design parameters such as gap width, loop area, pad dimensions, or substrate thickness, evaluated through FEM parameter sweeps.

Inverse-design logic therefore prioritizes those parameters with the largest \(\mathcal{S}_{G,j}\), namely the geometric changes that most efficiently reduce the coupling functional for the currently budget-limiting channel.

The resulting co-optimization tension is concrete. Enlarging a capacitor pad may reduce electric-field concentration at edges and thereby lower \(G_I\), while simultaneously increasing metal--air surface area and therefore increasing \(G_{\Phi}\). As a device-specific illustrative case from a recent design iteration, a \(30\%\) pad enlargement can reduce the Prescriptor I coupling functional by roughly \(15\%\) while increasing the Prescriptor II coupling functional by roughly \(8\%\), though the exact numbers remain geometry dependent.

The prescriptor formalism makes this trade-off quantitative and visible before fabrication, through independent FEM computation of both \(G_I\) and\(G_{\Phi}\) at the proposed geometry. Table~\ref{tab:B2_conflict} summarizes the qualitative conflict structure for several geometry-sensitive channels.

\subsection*{V. Validity Boundaries}

Inverse design is valid only within the separability regime. Four conditions must hold.

\begin{enumerate}
\item \textbf{No geometry-induced microstructural redistribution.} The geometric optimization must not redistribute the defect population. If FIB-SEM witness data from the modified geometry show a significant shift in the curvature distribution, the inverse calculation at that design point is unreliable.

\item \textbf{No process-induced change in \(G_j\).} A surface treatment intended to reduce \(\rho_j\) must not simultaneously alter the physical geometry in a way that changes \(G_j\). If it does, the change in \(G_j\) must be independently quantified.

\item \textbf{Moderate perturbation regime.} The inverse-design approach is most reliable for moderate geometric variations around a validated baseline. The 2\(\times\)2 protocol of Sec.~\ref{sec:falsifiability} of the main text provides the experimental diagnostic for identifying when this regime is exceeded.

\item \textbf{No intrinsic \(\rho\)--\(G\) coupling.} The framework breaks down if the microstructural state variable \(\rho_j\) and the coupling functional \(G_j\) are intrinsically coupled by the same physical mechanism. For example, in a strained thin-film system where lattice strain simultaneously modifies the defect density (shifting \(\rho_I\)) and the local field geometry (shifting \(G_I\) through altered dielectric properties), the product form ceases to represent two independently controllable degrees of freedom. In such cases, the 2\(\times\)2 protocol will detect the coupling through inconsistent row and column ratios, but the inverse-design targets derived from the separable product will not be reliable.
\end{enumerate}

In such cases, the factorized inverse targets are not reliable, even if they remain suggestive at leading order.

\subsection*{VI. Toward Quantitative Validation}

The inverse-design predictions outlined here are testable against published data. Process-split studies containing both curvature and coherence data can be used to compare extracted \(\mu_2\) values against measured \(T_1\). Shielding studies provide paired data for testing Prescriptor IV-style inverse predictions.

Validation can begin at the level of order-of-magnitude agreement between predicted and observed channel-wise loss rates. Exact numerical agreement is not expected at leading order, but gross inconsistency would count as falsification.

The inverse-design capability follows directly from the prescriptor factorization and requires no additional conceptual machinery. By converting coherence targets into channel-wise upper bounds on \(\rho_j\) and into geometric sensitivities \(G_j\), the framework provides a structured co-design interface between materials scientists, who control \(\rho_j\), and device engineers, who control \(G_j\).

\end{document}